\documentclass[a4paper,11pt]{article}
\pdfoutput=1 % if your are submitting a pdflatex (i.e. if you have
\usepackage{jheppub} % for details on the use of the package, please
\usepackage[T1]{fontenc} % if needed

\usepackage{amsmath,amssymb,epsfig,ulem,color,slashed,wasysym}
\allowdisplaybreaks 

\def \beq{\begin{equation}}
\def \eeq{\end{equation}}
\def \bea{\begin{eqnarray}}
\def \eea{\end{eqnarray}}

\title{\boldmath Anomalous triple gauge  couplings \\ from $B$-meson and kaon observables}

\author[1]{Christoph Bobeth}
\author[2,3]{and Ulrich Haisch}

\affiliation[1]{Technische Universit\"at M\"unchen, Institute for Advanced Study, \\ 
Lichtenbergstra{\ss}e 2a, D-85748 Garching, Germany}

\affiliation[2]{Rudolf Peierls Centre for Theoretical Physics,
    University of Oxford, \\ OX1 3NP Oxford, United Kingdom}
    
\affiliation[3]{CERN, Theory Division, \\ CH-1211 Geneva 23, Switzerland}

\emailAdd{christoph.bobeth@ph.tum.de}

\emailAdd{u.haisch1@physics.ox.ac.uk}

\abstract{
We consider the three CP-conserving dimension-6 operators that encode the leading new-physics effects  in the  triple gauge couplings. The contributions to the standard-model electromagnetic dipole and semi-leptonic vector and axial-vector interactions that arise from the insertions of these operators are calculated. We show that  radiative and rare $B$-meson decays provide, under certain assumptions, constraints on two out of the three anomalous couplings that are competitive with the restrictions obtained from LEP~II, Tevatron and LHC data. The constraints arising from the $Z \to b \bar b$ electroweak pseudo observables, $K \to \pi \nu \bar \nu$ and $\epsilon^\prime/\epsilon$ are also studied.
}

\preprint{FLAVOUR(267104)-ERC-93 \\ \hspace*{\fill} CERN-PH-TH/2015-041}

\begin{document} 

\maketitle

\flushbottom

\section{Introduction}
\label{sec:introduction}

The tremendous experimental success of the standard model (SM) suggests that, at the weak scale, fundamental interactions respect a $SU(3)_C \times SU(2)_L\times U(1)_Y$ gauge symmetry and that any new state that may exist in nature is heavier than all known elementary particles. The discovery of a boson with a mass of  $125 \, {\rm GeV}$ and measurements of its production and decay rates at the LHC, give strong evidence for the Higgs mechanism, in which  a linearly realised $SU(2)_L \times U(1)_Y$ symmetry is spontaneously broken to $U(1)_{\rm EM}$ by the Higgs vacuum expectation value.  Weak-scale  physics can therefore be described by an effective field theory (EFT) in which the SM Lagrangian is the leading term and new-physics effects are encoded in higher-dimensional operators constructed solely out of SM fields. Such an EFT description allows one to parametrise effects of heavy extra states in terms of operators with increasing dimension, which is equivalent to an expansion in inverse powers of a mass scale $\Lambda$ that characterises the onset of new dynamics.  

Constraints on the Wilson coefficients of the higher-dimensional operators come from electroweak (EW) precision observables. For instance, non-standard contributions to gauge boson two-point  functions are severely constrained by oblique parameters as measured at the $Z$ pole by LEP~I~\cite{ALEPH:2005ab}. Anomalous triple  gauge couplings~(TGCs) are bounded by measurements of gauge boson pair production performed at LEP~II, the Tevatron and LHC~(see~e.g.~\cite{Aaltonen:2009aa,Aaltonen:2012vu,Abazov:2012ze,Schael:2013ita,Corbett:2013pja,Dumont:2013wma,Chatrchyan:2013yaa,Chatrchyan:2013fya,Aad:2014mda,Ellis:2014jta,Falkowski:2014tna}). Higgs-boson properties are also affected by the same higher-dimensional operators, which implies that LHC  measurements of the Higgs production and decay rates provide another handle on anomalous TGCs.  Understanding the correlations between the different constraints and exploiting this knowledge to refine future  new-physics searches will be an important task in the coming years. 

In this article we extend the existing model-independent studies of the SM-EFT by deriving  constraints on the CP-conserving TGCs using quark-flavour observables. While such restrictions have been  studied previously \cite{Chia:1989gh, Peterson:1991pk, Numata:1991xy, Rizzo:1993zr, He:1993uj, Baillie:1993cx, Dawson:1993in, He:1993hx, He:1994dt, Alam:1997nk, Burdman:1998ie, He:1999kta}, there are (at least) two good reasons that call for a new analysis. First,  in the literature there exists no consensus on the analytic form of the flavour-changing $\gamma d_i \bar d_j$ and $Z d_i \bar d_j$  one-loop vertices that result from the anomalous TGCs. By recalculating the one-loop contributions, we identify which of the available results are correct and which are not. Second, all earlier analyses are by now obsolete because they all predate the precision measurements of radiative and rare $B$-meson decays by  BaBar, Belle and LHCb. Our study is based on the  assumption that the TGCs are the only couplings that receive non-zero initial conditions at  the scale $\Lambda$ where new physics enters. This  in particular means that we do neither allow for significant flavour-changing down-type  interactions nor modifications of the $Zb\bar b$  couplings to occur at the new-physics scale~$\Lambda$. Under this assumption, we show that radiative and rare $B$ and $K$ decays such as $B \to X_s \gamma$, $B_s \to \mu^+ \mu^-$, $B \to X_s \ell^+ \ell^-$, $B \to K^{(\ast)} \mu^+ \mu^-$, $B_s \to \phi \mu^+ \mu^-$, $K \to \pi \nu \bar \nu$ and $\epsilon^\prime/\epsilon$ can provide constraints on two of the three TGCs that are comparable with the restrictions arising from high-energy observations. Like in case of anomalous $Z t \bar t$ couplings, this finding illustrates the complementarity and synergy between indirect  \cite{Brod:2014hsa} and direct~\cite{Rontsch:2014cca} searches for physics beyond the SM.

The outline of this article is as follows. In Section~\ref{sec:preliminaries} we discuss the three effective interactions to be examined in this paper and review how the SM-EFT operators are related to the TGC Lagrangian. The explicit calculations of the effects of anomalous TGCs in quark-flavour physics and in the $Z \to b \bar b$ pseudo observables are presented in Section~\ref{sec:results}. Our numerical analysis is performed in Section~\ref{sec:numerics}, while Section~\ref{sec:conclusions} contains our conclusions. 

\section{Preliminaries}
\label{sec:preliminaries}

The primary goal of our work is to derive indirect constraints on the following effective Lagrangian 
\beq \label{eq:Leff} 
{\cal L}_{\rm TGC} = \sum_{i=\phi B, \phi W, 3W} \frac{C_i}{\Lambda^2} \, O_i \,,
\eeq
which contains three C and P invariant dimension-6 operators.  Like in~\cite{Hagiwara:1993ck}, we write these operators as
\beq \label{eq:Oi}
\begin{split}
O_{\phi B} & = \left ( D_\mu \phi \right)^\dagger \hat B^{\mu \nu} \left ( D_\nu \phi \right)   \,, \\[2mm]
O_{\phi W} & = \left ( D_\mu \phi \right)^\dagger \hat W^{\mu \nu} \left ( D_\nu \phi \right)   \,, \\[2mm]
O_{3W} & = {\rm Tr} \left ( \hat W_{\mu \nu} \hat W^{\nu \rho}  \hspace{0.25mm} {\hat {W}_\rho}^{\; \mu} \right ) \,,
\end{split}
\eeq
where $D_\mu \phi = \left ( \partial_\mu + i g^\prime B_\mu/2 +  i g \hspace{0.25mm} \sigma^a \hspace{0.25mm} W_\mu^a/2 \right) \phi$ is the covariant derivative acting on the Higgs doublet $\phi$ and $\sigma^a$ are the usual Pauli matrices. The hatted field strength tensors are defined as  $\hat B_{\mu \nu} = i g^\prime \left ( \partial_\mu B_ \nu - \partial_\nu B_\mu \right)/2$ and $\hat W_{\mu \nu} = i g \hspace{0.25mm} \sigma^a \left ( \partial_\mu W_\nu^ a - \partial_\nu W_\mu^ a  - g \hspace{0.25mm} \epsilon_{abc}   W_\mu^b  \hspace{0.25mm} W_\nu^c \right )/2$, while~$g^\prime$ and $g$ denotes the $U(1)_Y$ and $SU(2)_L$ gauge coupling, respectively.
The parameter~$\Lambda$ is the new-physics scale at which the effective operators $O_i$  are generated by integrating out heavy degrees of freedom. The Wilson coefficients $C_i$ in (\ref{eq:Leff}) are hence understood to be evaluated at $\Lambda$. The operators introduced above represent the  leading corrections to the anomalous~TGCs that do not contribute to gauge-boson propagators at tree level.\footnote{Besides $O_{\phi B}$, $O_{\phi W}$ and $O_{3W}$ there are other dimension-6 operators like for instance $O_{BW}=\phi^\dagger \hat B_{\mu \nu} \hat W^{\mu \nu} \phi$ that modify the TGCs. In contrast to (\ref{eq:Oi}),  $O_{BW}$ however induces a tree-level contribution to the oblique parameter~$S$. In turn its Wilson coefficient is already so tightly constrained by EW precision observables that it cannot give rise to relevant one-loop corrections in flavour physics. We thus do not consider operators of the type $O_{BW}$ in our article.  }

The $SU(2)_L \times U(1)_Y$ gauge-invariant operators introduced in (\ref{eq:Oi}) modify the  triple gauge-boson interaction vertices. We parametrise these interactions as  ($V=\gamma, Z$)~\cite{Hagiwara:1986vm} 
\beq \label{eq:LWWV}
\begin{split}
{\cal L}_{WWV} = -i g_{WWV} \, \bigg  [ & \hspace{0.25mm} g_1^V \left ( W_{\mu \nu}^+  \hspace{0.25mm} W^{-  \hspace{0.25mm} \mu}  \hspace{0.25mm} V^\nu - W_\mu^+  \hspace{0.25mm} V_\nu  \hspace{0.25mm} W^{-  \hspace{0.25mm} \mu \nu}  \right ) \\[2mm] & + \kappa_V  \hspace{0.25mm}  W_\mu^+  \hspace{0.25mm}  W_\nu^-  \hspace{0.25mm}  V^{\mu \nu} + \frac{\lambda_V}{m_W^2} \, W_{\mu \nu}^+ \hspace{0.25mm} W^{- \hspace{0.25mm} \nu \rho} \hspace{0.25mm} {V_\rho}^\mu \hspace{0.25mm} \bigg ] \,.
\end{split}
\eeq
The overall coupling strengths are defined by $g_{WW\gamma} = g s_w = g^\prime c_w= e$ and $g_{WWZ} = g c_w$ with $s_w$~($c_w$) denoting the sine (cosine) of the weak mixing angle, whereas $W_{\mu \nu}^{\pm} = \partial_\mu W_\nu^\pm - \partial_\nu W_\mu^\pm$ and $V_{\mu \nu} =   \partial_\mu V_\nu - \partial_\nu V_\mu$ with $W_\mu^\pm$ and $V_\mu$ referring to the physical gauge boson fields. 

Defining $g_1^V = 1 + \Delta g_1^V$ and $\kappa_V = 1 + \Delta \kappa_V$, the shifts $\Delta g_1^V$, $\Delta \kappa_V$ and the couplings $\lambda_V$ can be identified with the Wilson coefficients $C_i$. Explicitly, the following simple relations are obtained~\cite{Hagiwara:1986vm}
\beq  \label{eq:TGCs}
\Delta g_1^Z = \frac{m_Z^2}{2 \Lambda^2} \, C_{\phi W} \,, \qquad 
\Delta \kappa_\gamma = \frac{m_W^2}{2 \Lambda^2} \left ( C_{\phi B} + C_{\phi W} \right ) \,, \qquad 
\lambda_\gamma = \frac{3 g^2 m_W^2}{2 \Lambda^2} \, C_{3W} \,, 
\eeq
with $m_W$ ($m_Z$) denoting the mass of the $W$ ($Z$) boson. Furthermore, one has $ \Delta g_1^\gamma = 0$ as a result of gauge invariance and\footnote{These two relations are only  valid if the  analysis is limited to dimension-6 operators \cite{Hagiwara:1993ck}. For new-physics scales~$\Lambda$ sufficiently above the weak scale, dimension-8 terms  are however small, rendering them phenomenologically irrelevant at present.} 
\beq \label{eq:tworelations}
\Delta \kappa_Z = \Delta g_1^Z - \frac{s_w^2}{c_w^2} \, \Delta \kappa_\gamma \,, \qquad \lambda_Z = \lambda_\gamma \,.
\eeq
It follows that only three of the five non-zero anomalous TGCs are linearly independent. Hereafter we choose these three independent parameters to be $\Delta g_1^Z$, $\Delta \kappa_\gamma$ and $\lambda_\gamma$.  Note that the matching conditions (\ref{eq:TGCs}) hold after EW symmetry breaking and thus the Wilson coefficient $C_i$ appearing in these expressions should be evaluated at a scale  of  order~$m_W$.
 
It is important to realise that the effective Lagrangian introduced in (\ref{eq:Leff}) provides only a small subset   of the dimension-6 operators of the full SM-EFT Lagrangian~(cf.~\cite{Buchmuller:1985jz,Grzadkowski:2010es}).  In order to derive bounds on the anomalous TGCs (\ref{eq:TGCs}), we make in the following  the assumption that the Wilson coefficients of all the (relevant) SM-EFT operators but $C_{\phi B}$, $C_{\phi W}$ and $C_{3 W}$ vanish at the scale $\Lambda$. 

\section{Analytic results}
\label{sec:results}

The Lagrangian introduced in (\ref{eq:LWWV}) gives rise to contributions to radiative and rare $B$ decays, kaon physics and the $Z \to b \bar b$ decay.  In this section, we calculate the leading TGC corrections  to the phenomenologically most relevant of these observables.   The computation proceeds in two steps. First, renormalisation group (RG) evolution of the Wilson coefficients of the SM-EFT operators from the new-physics scale $\Lambda$ down to the weak  scale of  order~$m_W$. This step depends crucially on the choice of initial conditions, and we assume that only the TGC operators $O_{\phi B}$, $O_{\phi W}$ and $O_{3 W}$ receive non-zero Wilson coefficients at $\Lambda$. Second, calculation of matching corrections to the Fermi theory at the weak scale obtained by integrating out the top quark as well as the Higgs, $Z$ and $W$ boson.

In the case of the $b \to s \gamma, \ell^+ \ell^-$ observables, one determines in this way the corrections~$\Delta C_i$ to the Wilson coefficients $C_i = (C_i)_{\rm SM} + \Delta C_i$ of the operators that enter the effective Lagrangian of $|\Delta B|=|\Delta S|=1$ decays
\beq \label{eq:Heff} 
{\cal L}_{|\Delta B|=|\Delta S|=1} =  \frac{4 G_F}{\sqrt{2}} \, V_{ts}^\ast V_{tb} \, \big  ( C_7  \hspace{0.25mm} Q_7 + C_9 \hspace{0.25mm} Q_9 + C_{10} \hspace{0.25mm} Q_{10} \big ) + {\rm h.c.} 
\eeq
Here and below the  $|\Delta B| = |\Delta S| = 1$ Wilson coefficients $C_i$ as well as $\Delta C_i$ ($i = 7,9,10$) are understood to be evaluated at $m_W$,  the Fermi constant is denoted by $G_F$ and  $V_{ij}$ are the elements of the Cabibbo-Kobayashi-Maskawa matrix.  The electromagnetic dipole, semi-leptonic vector and axial-vector operators are defined as 
\beq \label{eq:Q710}
\begin{split}
Q_7 & = \frac{e}{(4 \pi)^2} \, m_b \left ( \bar s_L  \hspace{0.25mm} \sigma_{\alpha \beta}  \hspace{0.25mm} b_R \right ) F^{\alpha \beta} \,, \\[2mm]
Q_9 & =  \frac{e^2}{(4 \pi)^2}  \left ( \bar s_L  \hspace{0.25mm} \gamma_\alpha  \hspace{0.25mm} b_L \right ) \left (\bar \ell  \hspace{0.25mm}  \gamma^\alpha  \hspace{0.25mm}  \ell \right ) \,, \\[2mm]
Q_{10} & = \frac{e^2}{(4 \pi)^2}  \left ( \bar s_L  \hspace{0.25mm} \gamma_\alpha  \hspace{0.25mm} b_L \right ) \left (\bar \ell  \hspace{0.25mm}  \gamma^\alpha  \gamma_5 \hspace{0.25mm}  \ell \right )  \,,
\end{split}
\eeq
with $L$, $R$ indicating the chirality of the fermionic fields, $\sigma_{\alpha \beta} = i \left [ \gamma_\alpha, \gamma_\beta \right ]/2$ and $F_{\alpha \beta}$ the field strength tensor of QED. The  effective Lagrangians for $b \to d$ and $s \to d$ transitions are found from (\ref{eq:Heff}) and (\ref{eq:Q710}) by  appropriate replacements.  The current-current and QCD penguin operators are not affected by the TGCs at the one-loop level. 

\begin{figure}[!t]
\begin{center}
\includegraphics[width=0.675 \textwidth]{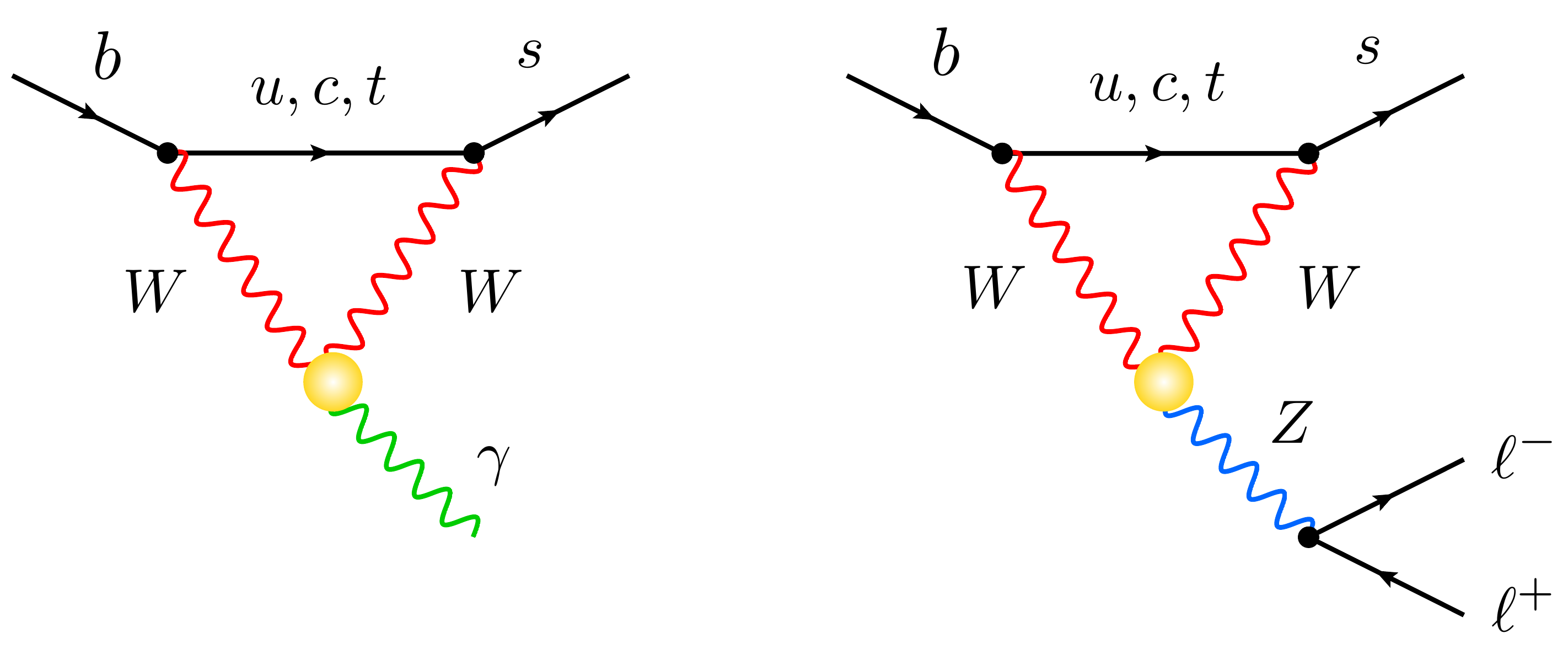} 
\end{center}
\vspace{-4mm}
\caption{\label{fig:diagrams} Examples of one-loop diagrams that generate a $b \to s \gamma$ (left) and a $b \to s \ell^+ \ell^-$~(right) transition.  The  insertions of the TGC operators are indicated by yellow circles, while the SM vertices are represented by black dots.}
\end{figure}

The Feynman graphs that give rise to the modifications $\Delta C_7$, $\Delta C_9$ and $\Delta C_{10}$ are depicted in Figure~\ref{fig:diagrams}.  Note that in the diagrams the contributions of all up-type quarks have to be considered in order to obtain the correct $b \to s \gamma$ and $b \to s \ell^+ \ell^-$ amplitudes. In terms of the three parameters $\Delta g_1^Z$, $\Delta \kappa_\gamma$ and $\lambda_\gamma$~$\big($see~(\ref{eq:TGCs})$\big)$, we arrive at 
\beq \label{eq:DeltaC710calculation}
\begin{split}
\Delta C_7 & = \Delta \kappa_\gamma \hspace{0.25mm} f_7^\kappa (x_t) + \lambda_\gamma  \hspace{0.25mm} f_7^\lambda (x_t) \,,  \\[2mm]
\Delta C_9 & = -\Delta D + \lambda_\gamma  \hspace{0.25mm} f_9^\lambda (x_t) + \frac{1-4s_w^2}{s_w^2} \, \Delta C \,, \\[2mm]
\Delta C_{10} & = -\frac{1}{s_w^2} \, \Delta C \,,
\end{split}
\eeq
with $x_t = m_t^2/m_W^2$ and 
\beq \label{eq:f79}
\begin{split}
f_7^\kappa (x) & = -\frac{x}{2 \left (x-1 \right )^2} -  \frac{x^3-3 x^2}{4 \left (x-1 \right )^3} \, \ln x \,, \\[2mm]
f_7^\lambda (x) & = \frac{x^2+x}{4 \left (x-1 \right )^2}-\frac{x^2}{2 \left (x-1 \right )^3} \, \ln x \,, \\[2mm]
f_9^\lambda (x) & = -\frac{3 x^2-x}{2 \left (x-1 \right )^2} +\frac{x^3}{\left (x-1 \right )^3} \, \ln x \,.
\end{split}
\eeq
For the universal $Z$-penguin (photon-penguin) coefficient $\Delta C$ ($\Delta D$) entering  both $\Delta C_9$ and $\Delta C_{10}$ ($\Delta C_9$), we obtain 
\beq \label{eq:DeltaCD} 
\Delta C  = \frac{3 c_w^2}{8} \, \Delta g_1^Z \,  x_{t} \, \ln \left ( \frac{\Lambda^2}{m_W^2} \right ) \,, \qquad
\Delta D  = \frac{1}{4} \, \Delta \kappa_\gamma \,  x_{t} \, \ln \left ( \frac{\Lambda^2}{m_W^2} \right ) \,.
\eeq
Our analytic result for $\Delta C_7$ as given in (\ref{eq:DeltaC710calculation}) agrees with the expression that derives from the calculations \cite{Chia:1989gh, Numata:1991xy, He:1993hx}. It however differs from the findings reported in~\cite{Peterson:1991pk,Rizzo:1993zr,Burdman:1998ie}.  In the case of $\Delta C_9$ and $\Delta C_{10}$,  the given expressions  can be obtained by combining the results of~\cite{Peterson:1991pk, Baillie:1993cx, He:1993uj, He:1999kta}. On the other hand, our result for the term in $\Delta C_9$ proportional to $\Delta \kappa_\gamma$ does not agree with  \cite{Chia:1989gh}, while it differs by a sign from~\cite{He:1994dt}. 

The Inami-Lim functions~(\ref{eq:f79}) and~(\ref{eq:DeltaCD}) can be divided into two categories. First, the large logarithms in~(\ref{eq:DeltaCD}) encode the leading-logarithmic (LL) contributions associated to the RG evolution between $\Lambda$ and $m_W$. We have verified that the coefficients of these~$\ln \left ( \Lambda^2/m_W^2 \right)$ terms can be recovered from the anomalous dimensions given in~\cite{Jenkins:2013wua}. This  provides a non-trivial cross-check since the latter calculation is performed in a  different operator basis \cite{Buchmuller:1985jz,Grzadkowski:2010es}.  It furthermore shows that the corresponding ultraviolet~(UV) divergences are properly absorbed into  counterterms in the context of the complete operator basis of the SM-EFT. Notice that a resummation of  LL effects  is possible, but given that $y_t^2/(4\pi)^2 \hspace{0.25mm} \ln \left ( \Lambda^2/m_W^2 \right ) \ll 1$ with $y_t$ the top-quark Yukawa coupling, keeping the $\ln \left (\Lambda^2/m_W^2 \right)$ terms unresummed is a good approximation. A second type of contributions to the coefficients $\Delta C_i$ comes from the evaluation of the matrix elements of the TGC operators at the scale~$m_W$. The corrections of this sort are encoded in the Inami-Lim functions (\ref{eq:f79}) and do not contain $\ln \left (\Lambda^2/m_W^2 \right)$ terms. They constitute the leading effects of  $\Delta\kappa_\gamma$ in $C_7$ and $\lambda_\gamma$ in $C_7$ and $C_9$. On the other hand, finite matching corrections are not included in the case of the LL universal terms  (see for instance~\cite{Freitas:2008vh, Grzadkowski:2008mf} for related  discussions) since they are formally subleading,~i.e.~of~next-to-leading logarithmic~(NLL)~order. Finally, also the evolution of the Wilson coefficients $C_i$ from $\Lambda$ down to~$m_W$ associated  to ``self-mixing'' has been systematically neglected when making use of~(\ref{eq:TGCs}) to  rewrite~$C_i(\Lambda)$ in terms of the TGCs, as it constitutes a NLL effect as well. We add that this self-mixing, $C_i(\Lambda) = \left [\delta_{ij} + \gamma_{ij}/(4 \pi)^2 \ln \left (\Lambda^2/m_W^2 \right) \right ] C_j(m_W)$, does not involve top-quark Yukawa enhanced anomalous dimensions $\gamma_{ij}$, and therefore is not only formally suppressed, but in practice  has  also a minor numerical impact. Incorporating beyond-LL corrections might however be mandatory in future precision studies.

We finally observe that the functions $f_7^\kappa (x)$ and $f_7^\lambda (x)$ in (\ref{eq:f79}) scale as $x$ for $x\ll1$. This implies that the TGC corrections to the anomalous magnetic moment of the muon $a_\mu$ are proportional to $\sum_{i=1}^3 |U_{\mu i}|^2 \, m_{\nu_i}^2/m_W^2$ with $U_{\mu i}$ denoting the elements of the Pontecorvo-Maki-Nakagawa-Sakata mixing matrix.  Due to the strong chiral suppression by the neutrino masses, the TGC contributions to $a_\mu$ are unobservably small. The same statements  apply  to lepton-flavour violating decays like $\mu \to e \gamma$ and $\mu \to 3 e$.

The coefficient $\Delta C$ also contributes to the rare $K \to \pi \nu \bar \nu$ decays, while both $\Delta C$ and~$\Delta D$ affect the observable $\epsilon^\prime/\epsilon$, which measure the amount of direct CP violation in the $K \to \pi \pi$ system. It is well-known (see e.g.~Section III.2.2 of \cite{Kronfeld:2013uoa}) that these observables are very sensitive probes of modifications in the EW penguin sector, and we will give the relevant formulas that allow for a phenomenological analysis in the next section. 

In the presence of (\ref{eq:TGCs}) the interactions between the $Z$ boson and bottom quarks are modified at the one-loop level as well. We parameterise the resulting effective $Z b \bar b$ couplings at the weak scale as follows 
\beq \label{eq:LZbb}
{\cal L}_{Zb\bar b} = \frac{e}{s_w c_w} \left [ \left ( g_L^b +\delta g_L^b \right ) \bar b_L \hspace{0.25mm} \slashed{Z} \hspace{0.25mm} b_L +  g_R^b \hspace{0.5mm} \bar b_R \hspace{0.25mm} \slashed{Z} \hspace{0.25mm} b_R  \right ] \,,
\eeq
with $g_L^b = -1/2 + s_w^2/3$ and $g_R^b = s_w^2/3$. For the one-loop correction $\delta g_L^b$ associated  to the anomalous TGCs, we obtain the compact expression 
\beq \label{eq:dgLb}
\delta g_L^b = \frac{\alpha}{2 \pi s_w^2} \, |V_{tb}|^2 \, \big (  \Delta C +  \delta C \big )\,, 
\eeq 
where $\Delta C$ can be found in (\ref{eq:DeltaCD}) and 
\beq \label{eq:deltaC}
\delta C = \frac{1}{16} \, \Delta \kappa_Z \, x_t  \, \ln \left ( \frac{\Lambda^2}{m_W^2} \right ) \,,
\eeq
with $\Delta \kappa_Z$ given in (\ref{eq:tworelations}). Note that the correction $\delta C$ in (\ref{eq:dgLb}) arises from expanding the $Z\to b \bar b$ amplitude to second order  with respect to the external momenta (cf.~\cite{Haisch:2007ia,Haisch:2011up}). At the $Z$ pole these effects scale as $m_Z^2/m_W^2$ with respect to $\Delta C$, while in $B$-meson (kaon) decays they are of order $m_b^2/m_W^2$~($m_c^2/m_W^2$), and thus can be safely neglected. Like in the case of $\Delta D$ and $\Delta  C$, we have included in $\delta C$ only the LL effects. Our results (\ref{eq:dgLb}) and~(\ref{eq:deltaC}) agree with the  expressions given in \cite{Alam:1997nk}.   

\section{Numerical analysis}
\label{sec:numerics}

In this section we analyse the constraints on $\Delta g_1^Z$, $\Delta \kappa_\gamma$ and $\lambda_\gamma$ that arise from indirect probes. We start our discussion by considering the inclusive radiative $B$-meson decay. In terms of the shift $\Delta C_7$, the ratio between the new-physics branching ratio of $B \to X_s \gamma$ and the SM prediction can be written as \cite{Misiak:2015xwa}
\beq \label{eq:RXs}
R_{X_s} = \frac{{\rm Br} \left (B \to X_s \gamma \right )}{{\rm Br} \left (B \to X_s \gamma \right )_{\rm SM}} = 1 -2.45  \hspace{0.25mm} \Delta C_7  \,.
\eeq
Combining the latest SM calculation \cite{Misiak:2015xwa,Czakon:2015exa} with the present world average~\cite{Amhis:2014hma}, one gets 
\beq \label{eq:RXsexp}
R_{X_s} = 1.02 \pm 0.10 \,,
\eeq
if uncertainties are added in quadrature. 

In the case of the $B_s \to \mu^+ \mu^-$ decay, we arrive instead at the simple formula 
\beq \label{eq:Rmm} 
R_{\mu^+ \mu^-} = \frac{{\rm Br} \left (B_s \to \mu^+ \mu^- \right )}{{\rm Br} \left (B_s \to \mu^+ \mu^- \right )_{\rm SM}} = \big ( 1 - 0.25  \hspace{0.25mm} \Delta C_{10} \big )^2 \,.
\eeq
To constrain the coefficient $\Delta C_{10}$, this result should be compared to 
\beq \label{eq:Rmmexp}
R_{\mu^+ \mu^-} = 0.78 \pm 0.18 \,, 
\eeq
which we have obtained from the recently improved SM prediction \cite{Bobeth:2013uxa, Hermann:2013kca, Bobeth:2013tba} and the combined CMS and LHCb result~\cite{CMS:2014xfa} by treating the individual uncertainties as Gaussian errors.  

In order to set bounds on the TGCs  from  the  modified $Zb\bar b$ couplings~(\ref{eq:LZbb}), we employ the result of a recent global analysis~\cite{Ciuchini:2013pca} of the complete  set of 15 EW precision observables,  which leads to 
\beq \label{eq:deltagLbexp}
\delta g_L^b = 0.0016 \pm 0.0015 \,.
\eeq
The given central value and uncertainty is derived computing the region containing 68\% of a marginalised probability distribution function and symmetrising the error.

\begin{figure}[!t]
\begin{center}
\includegraphics[width=0.3 \textheight]{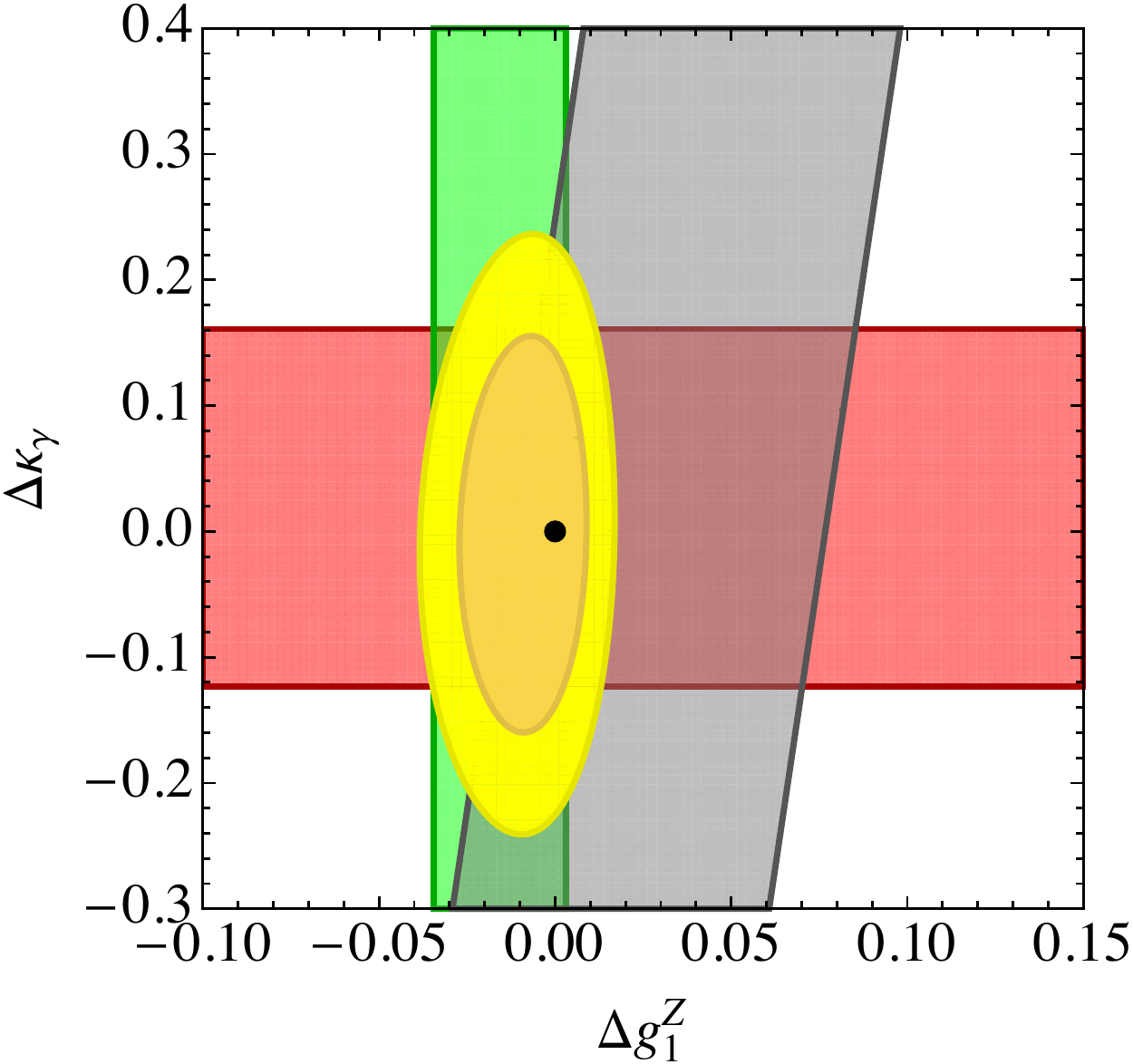}  \qquad 
\includegraphics[width=0.3 \textheight]{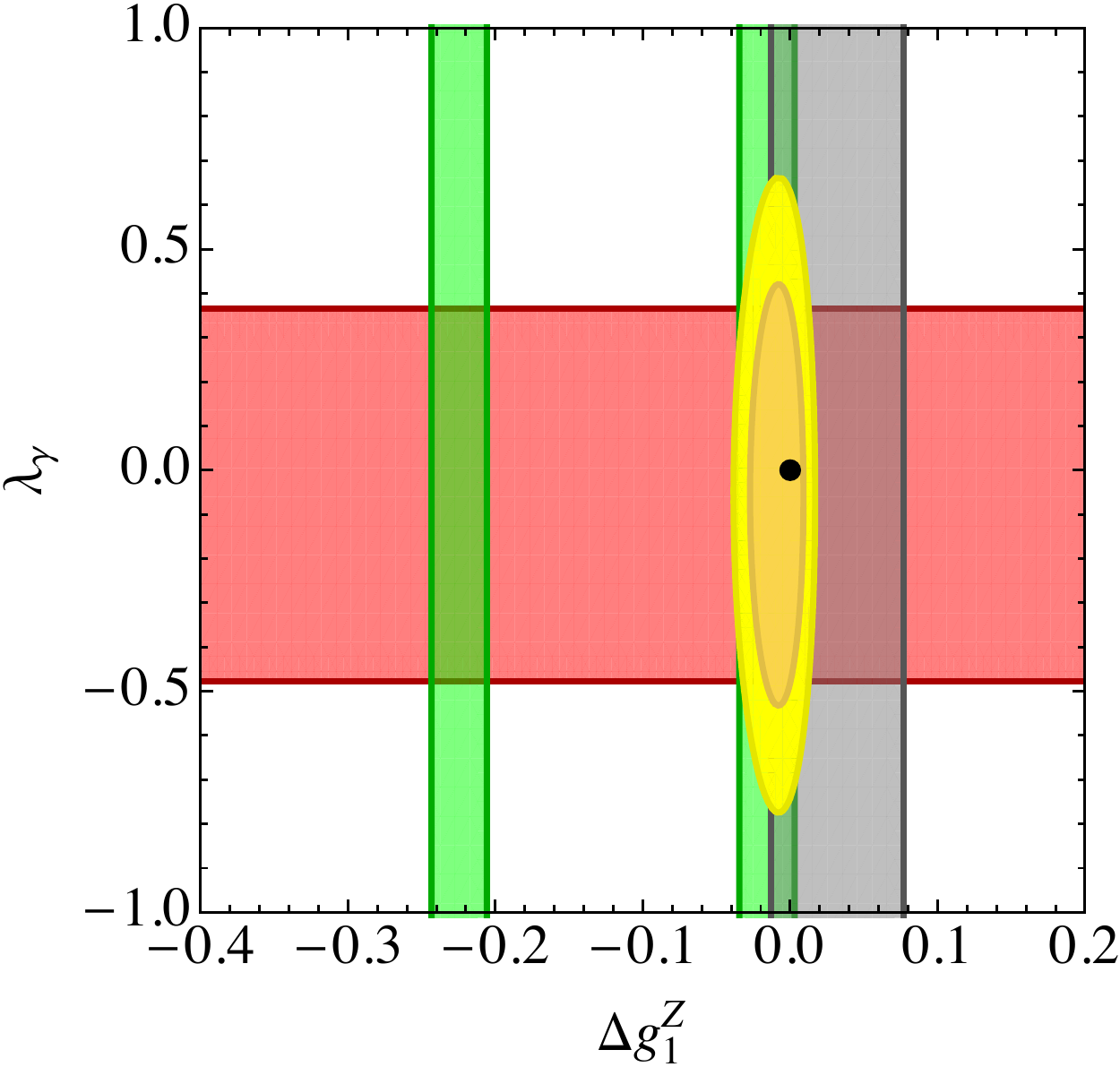} 

\vspace{2mm}

\includegraphics[width=0.3 \textheight]{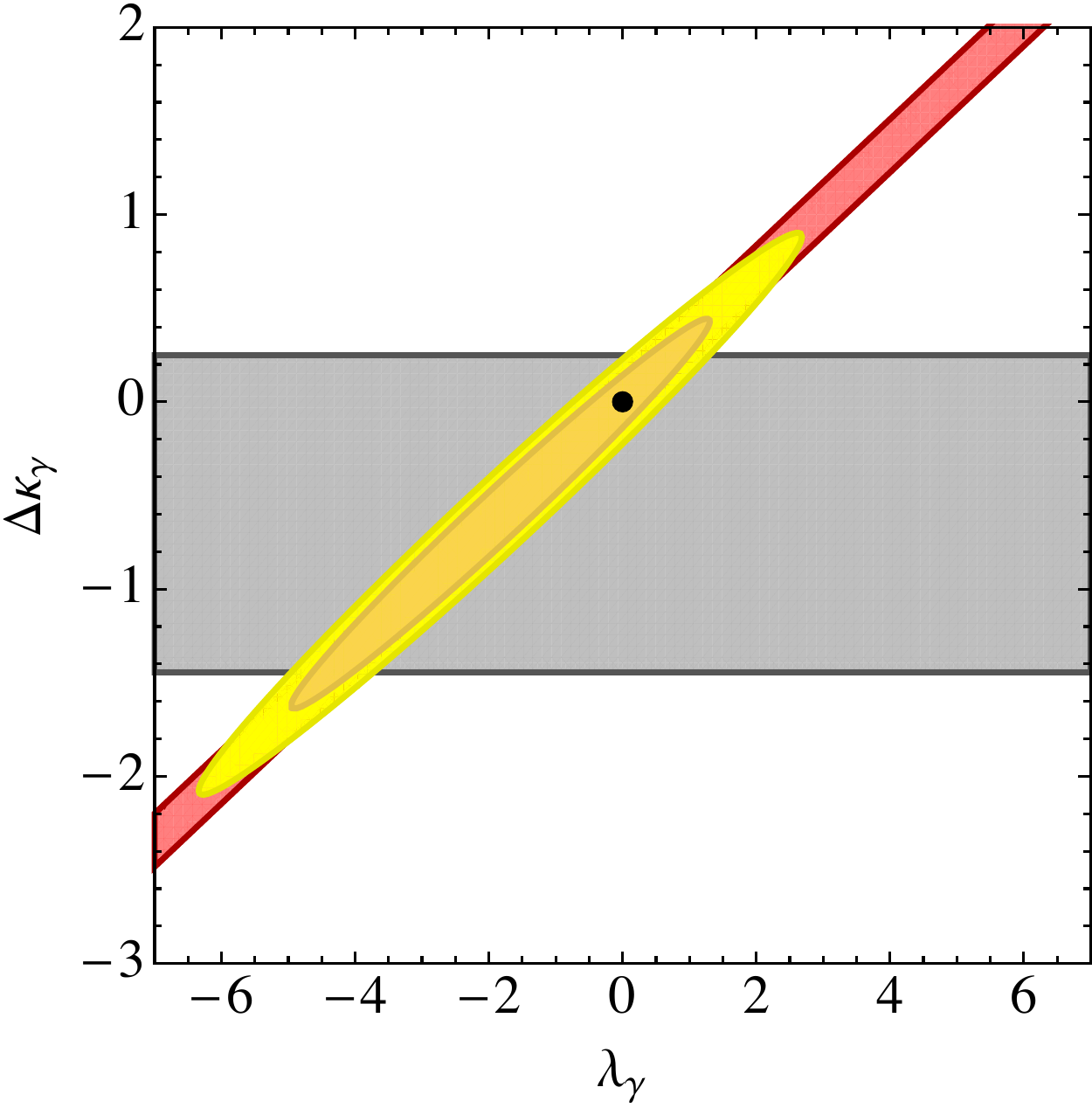} 
\end{center}
\vspace{-4mm}
\caption{\label{fig:panels1} Allowed regions in the $\Delta g_1^Z \hspace{0.25mm}$--$ \hspace{0.25mm} \Delta \kappa_\gamma$ plane (upper left panel),  the $\Delta g_1^Z \hspace{0.25mm}$--$ \hspace{0.25mm} \lambda_\gamma$ plane (upper right panel) and the $\lambda_\gamma  \hspace{0.25mm}$--$ \hspace{0.25mm} \Delta \kappa_\gamma$ plane (lower panel). The red, green and grey contours correspond to the 68\% CL best fit regions following from $B \to X_s \gamma$, $B_s \to \mu^+ \mu^-$ and $Z \to b \bar b$, respectively.  The yellow and orange contours show the 68\% CL and 95\% CL regions arising from a combination of the individual constraints.   The shown constraints have been obtained for $\Lambda = 2 \, {\rm TeV}$. The black points correspond to the SM.}
\end{figure}

It is important to realise that the SM predictions for ${\rm Br} \left (B \to X_s \gamma \right )$,  ${\rm Br} \left (B_s \to \mu^+ \mu^- \right )$ and $\Gamma \left (Z \to b \bar b \right )$ are all under good control, having theoretical uncertainties below the $10 \%$~level. Combining the three constraints $R_{X_s}$, $R_{\mu^+ \mu^-}$ and $\delta g_L^b$ therefore allows for a clean extraction of the TGC parameters $\Delta g_1^Z$, $\Delta \kappa_\gamma$ and $\lambda_\gamma$. The resulting constraints on the different pairs of coefficients are shown in the three panels of Figure~\ref{fig:panels1}. The  yellow and orange contours represent the 68\%~confidence level (CL)~($\Delta \chi^2 = \chi^2 - \chi_{\rm SM}^2 =2.3$ with  $\chi_{\rm SM}^2 = 2.6$) and 95\%~CL~($\Delta \chi^2 = 6.0$) regions arising from the global fit, while the coloured bands  correspond to the 68\% CL regions  following from the individual constraints. The parameter not depicted in each panel has been set to zero and the new-physics scale~$\Lambda$ appearing in~(\ref{eq:DeltaCD}) and (\ref{eq:deltaC}) has been fixed to~$2 \, {\rm TeV}$.  While this choice will be the benchmark throughout our article, we will also comment on how our numerical results change if lower values of~$\Lambda$ are employed.

From the panels it is evident that the most tightly constrained TGC is $\Delta g_1^Z$, and that these bounds are driven by the constraint (\ref{eq:Rmmexp}). This feature is readily understood by noticing that the relevant Wilson coefficient scales as $\Delta C_{10} \sim -\Delta g_1^Z \, x_t \ln \left (\Lambda^2/m_W^2 \right )$, whereas one has  $\Delta C_7 \sim -\Delta \kappa_\gamma \ln x_t$  in the  limit $x_t \to \infty$ of infinitely heavy top quark. The  strong sensitivity of $B_s \to \mu^+ \mu^-$ to the TGC $\Delta g_1^Z$, which by now exceeds the one of the $Z \to b \bar b$ pseudo observables, nicely illustrates the role of this decay mode as an EW precision test~\cite{Guadagnoli:2013mru}. Given that the $Z b \bar b$ couplings have been measured with a precision of around~$0.5\%$, while the branching ratio of $B_s \to \mu^+ \mu^-$ is only known to roughly~$25\%$, this finding may be surprising at first sight. The naive expectation is upset by the fact that in the case of $B_s \to \mu^+ \mu^-$ both the SM and the new-physics contributions arise at the one-loop level, while for $Z \to b \bar b$ the one-loop TGC corrections have to compete against a SM tree-level contribution.  This effectively boosts  the  sensitivity of $B_s \to \mu^+ \mu^-$  relative to that of the $Zb \bar b$ pseudo observables by a factor of about $\pi/\alpha_w \simeq 100$ with $\alpha_w \simeq 0.03$ the weak coupling constant.

Explicitly, we find in the case~$\lambda_\gamma = 0$~(upper left panel)  the following 68\%~CL constraints
\beq \label{eq:fit1}
\Delta g_1^Z = -0.009 \pm 0.019 \,, \qquad 
\Delta \kappa_\gamma = 0.00 \pm 0.16 \,, 
\eeq
and the correlation matrix 
\beq
\rho = \begin{pmatrix} 1 & -0.06 \\  -0.06 & 1 \end{pmatrix} \,.
\eeq
The above numbers  should be contrasted  with the limits that can be derived from EW gauge boson pair production at LEP~II, the Tevatron and the LHC as well as Higgs physics~(see~e.g.~\cite{Aaltonen:2009aa,Aaltonen:2012vu,Abazov:2012ze,Schael:2013ita,Corbett:2013pja,Chatrchyan:2013yaa,Chatrchyan:2013fya,Aad:2014mda,Ellis:2014jta,Falkowski:2014tna}). For instance, assuming $\left |\lambda_\gamma \right | \ll \left |\Delta g_1^Z \right |, \left |\Delta \kappa_\gamma \right |$, the recent analysis~\cite{Falkowski:2014tna} of the $e^+ e^- \to W^+ W^-$  production data collected by the LEP~II experiments  obtains   the model-independent 68\%~CL bounds $\Delta g_1^Z = -0.06 \pm 0.03$ and $\Delta \kappa_\gamma = 0.06 \pm 0.04$. We see that compared to (\ref{eq:fit1}) the LEP~II limit on $\Delta g_1^Z$ is slightly weaker, while in the case of the parameter $\Delta \kappa_\gamma$ the LEP~II constraint is clearly superior. We add that if instead of $\Lambda = 2 \, {\rm TeV}$ the value $\Lambda = 1 \, {\rm TeV}$ ($\Lambda = 0.5 \, {\rm TeV}$) is used in the combined fit~(\ref{eq:fit1}), one obtains $\Delta g_1^Z = -0.012 \pm 0.024$ and $\Delta \kappa_\gamma = 0.00 \pm 0.16$ ($\Delta g_1^Z = -0.016 \pm 0.033$ and $\Delta \kappa_\gamma = 0.01 \pm 0.16$). These numbers show that the bounds on $\Delta g_1^Z$ and $\Delta \kappa_\gamma$, when obtained from $B_s \to \mu^+ \mu^-$ and $B \to X_s \gamma$, are only weakly dependent on the precise value of the new-physics scale $\Lambda$.

From the upper right and lower panel in Figure~\ref{fig:panels1} one observes in addition that the constraints on $\lambda_\gamma$ are notably less stringent than those on $\Delta g_1^Z$ and $\Delta \kappa_\gamma$. The reason for this is twofold. First, only $B \to X_s \gamma$ is sensitive to $\lambda_\gamma$ and the contribution is not logarithmically enhanced, scaling like $\Delta C_7 \sim \lambda_\gamma$. Second, the $R_{X_s}$ constraint (\ref{eq:RXs}) has an approximately blind direction along $\Delta \kappa_\gamma \simeq 1/3  \, \lambda_\gamma$. In consequence, our bounds on  $\lambda_\gamma$ are not competitive with those that derive for instance from EW gauge boson production at LEP~II, which read~$\lambda_\gamma =  0.00 \pm 0.07$~\cite{Falkowski:2014tna}.  Observe  that the  $R_{\mu^+ \mu^-}$ constraint allows in principle for large destructive new-physics contributions that would effectively flip the sign of the $B_s \to \mu^+ \mu^-$ amplitude. This feature is illustrated by the second green band in the upper right panel. One also sees that the corresponding ambiguity in~$\Delta g_1^Z$ is resolved by  $Z \to b \bar b$. In the case of $B \to X_s \gamma$ a similar ambiguity exists, but  it is not shown in the lower panel, because the existing informations on $b \to s \ell^+ \ell^-$ transitions can be used to eliminate this possibility~\cite{Gambino:2004mv}.\footnote{The latest model-independent fit that allows for all solutions  finds a strong posterior preference for the SM-like sign solution of $C_{7}$, $C_9$ and $C_{10}$ of 4 to 1 compared to the sign-flipped case \cite{Beaujean:2013soa}.}

\begin{figure}[!t]
\begin{center}
\includegraphics[width=0.3 \textheight]{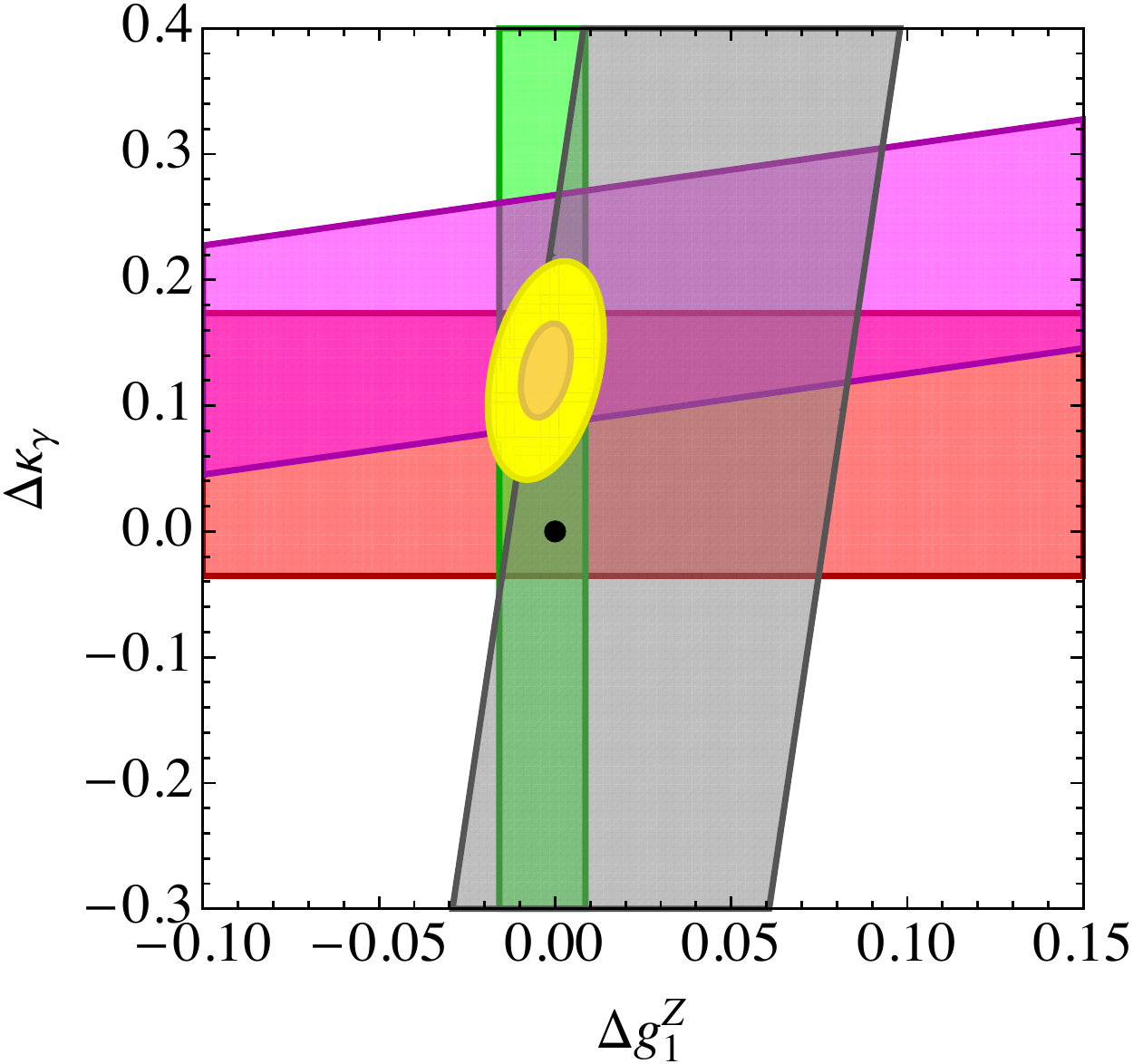}  \qquad 
\includegraphics[width=0.3 \textheight]{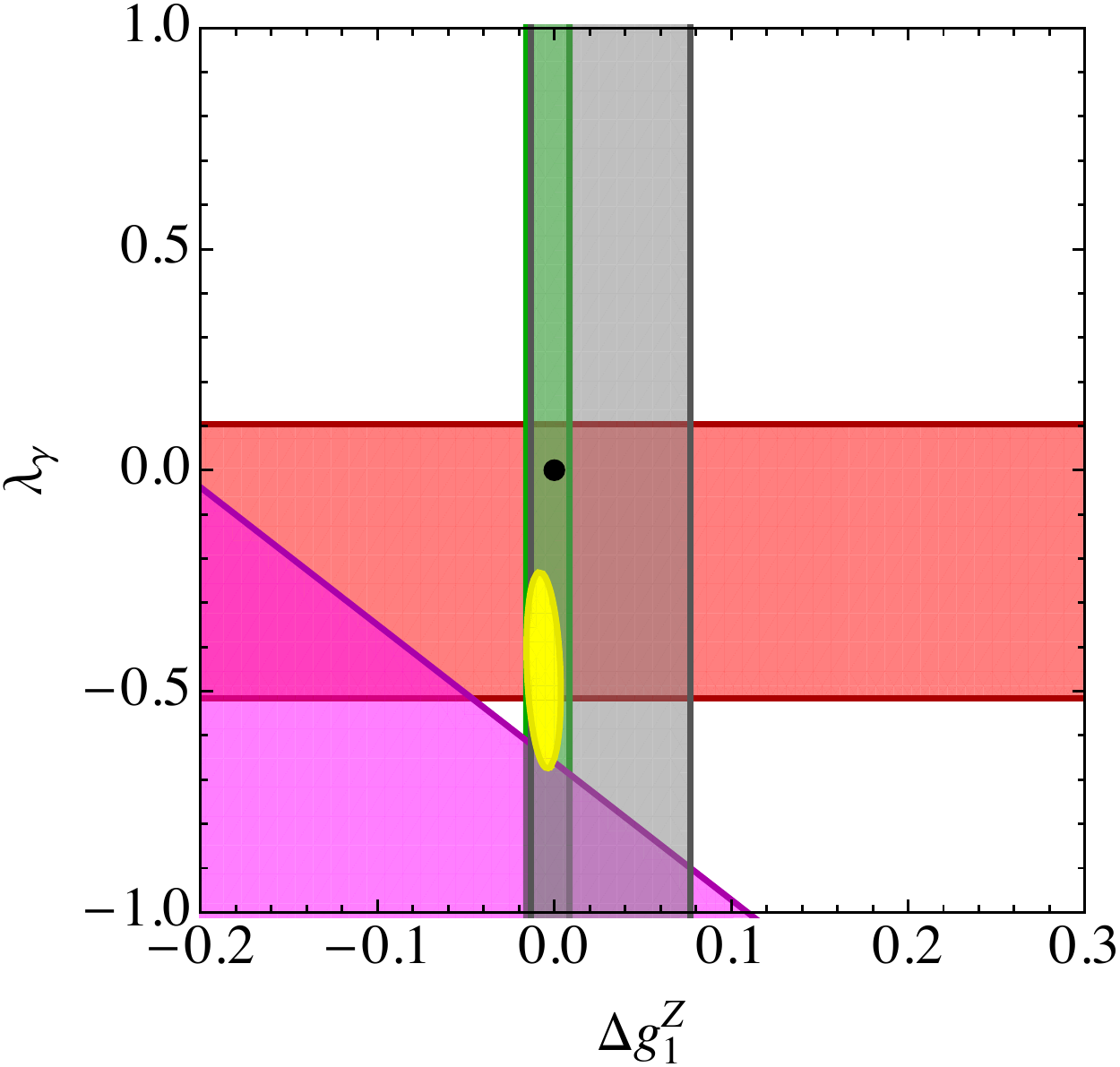} 

\vspace{2mm}

\includegraphics[width=0.3 \textheight]{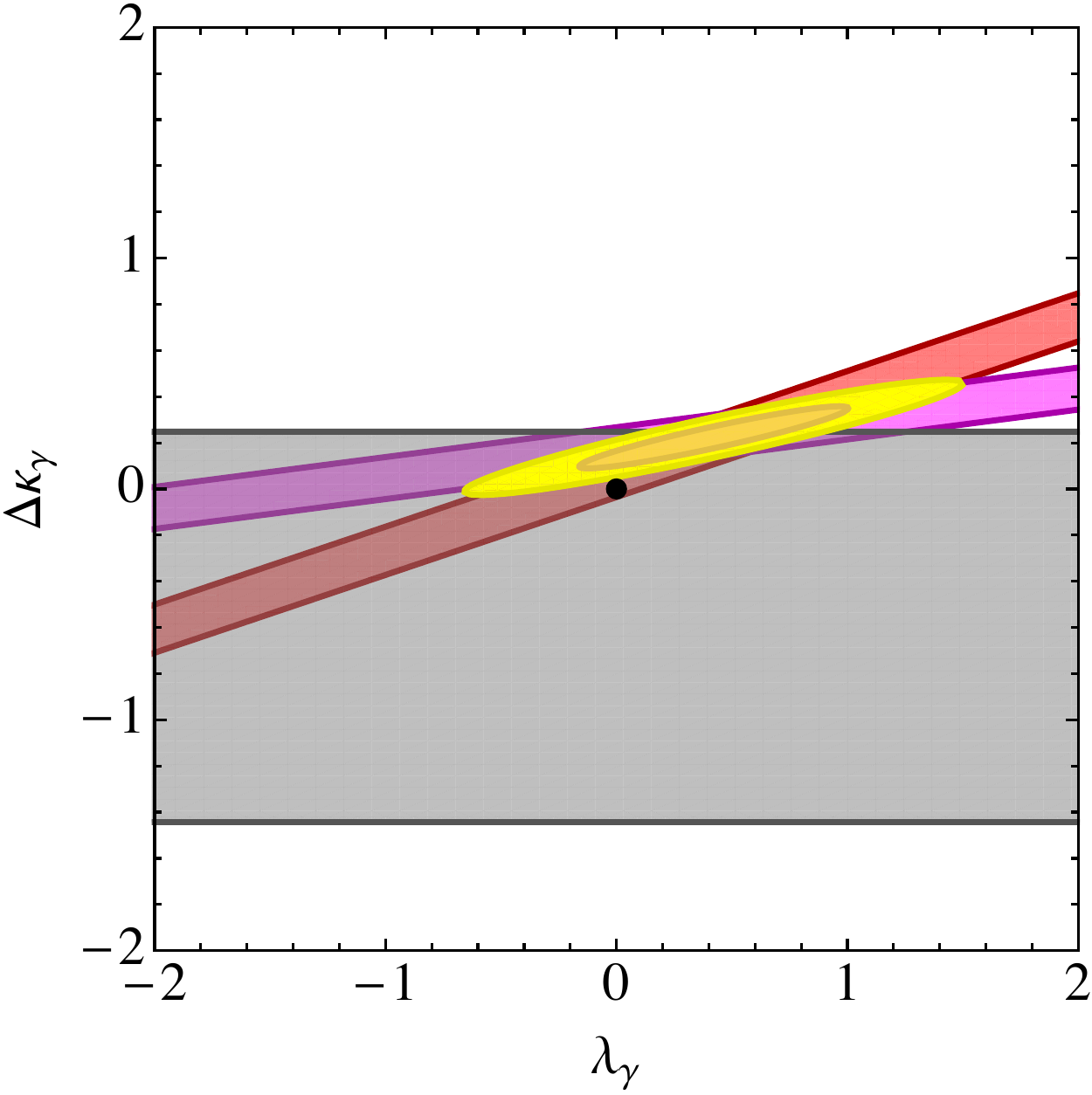} 
\end{center}
\vspace{-4mm}
\caption{\label{fig:panels2} Allowed regions in the $\Delta g_1^Z \hspace{0.25mm}$--$ \hspace{0.25mm} \Delta \kappa_\gamma$ plane (upper left panel),  the $\Delta g_1^Z \hspace{0.25mm}$--$ \hspace{0.25mm} \lambda_\gamma$ plane (upper right panel) and the $\lambda_\gamma  \hspace{0.25mm}$--$ \hspace{0.25mm} \Delta \kappa_\gamma$ plane (lower panel). The red, magenta, green and grey contours correspond to the 68\% CL best fit regions following from $\Delta C_7$, $\Delta C_9$, $\Delta C_{10}$ and $\delta g_L^b$, respectively. The dominant constraint on $\Delta C_7$ arises from $B \to X_s \gamma$, while $\Delta C_9$ is bounded most strongly by $B \to K^\ast \mu^+ \mu^-$. The constraint $\Delta C_{10}$ is driven by the combination of  $B \to K^{(\ast)} \mu^+ \mu^-$, $B_s \to \phi  \mu^+ \mu^-$ and $B_s \to \mu^+ \mu^-$. The yellow and orange contours show the 68\% CL and 95\% CL regions arising from a combination of the individual constraints.   All constraints employ $\Lambda = 2 \, {\rm TeV}$. The black points correspond to the SM.}
\end{figure}

The quark-flavour observables (\ref{eq:RXs}) and (\ref{eq:Rmm}) that we have considered so far are only sensitive to the shifts $\Delta C_7$ and $\Delta C_{10}$, but  are unaffected by a possible new-physics contribution~$\Delta C_9$. In order to gain sensitivity to $\Delta C_9$ one has to include informations on the $b \to s \ell^+ \ell^-$ transitions in the global fit. The recent analysis \cite{Altmannshofer:2014rta} is based on $78$ independent measurements of radiative  and rare $b \to s$ observables. It finds the   constraints
\beq \label{eq:DeltaC710}
\Delta C_7 = -0.03 \pm 0.03 \,, \qquad 
\Delta C_9 = -1.17 \pm 0.40 \,, \qquad 
\Delta C_{10} = 0.12 \pm 0.27 \,, 
\eeq
at 68\%~CL and the correlations 
\beq 
\rho = \begin{pmatrix} 1 & -0.22 & \ 0.02 \\
-0.22 & 1 & \  0.31 \\
0.02 & 0.31 & \  1
\end{pmatrix} \,.
\eeq
The corresponding $\chi^2$ in the SM is $104.6$, while the best-fit point of the new-physics hypothesis has a $\chi^2$ of $90.6$. Note that (\ref{eq:DeltaC710}) correspond to a tension between the SM predictions and the data at the level of $3 \sigma$.  While further LHCb studies of $B \to K^\ast \mu^+ \mu^-$, $B^+ \to K^+ \mu^+ \mu^-$ and $B_s \to \phi  \mu^+ \mu^-$ and a better understanding of hadronic uncertainties  will be necessary to clarify whether the observed deviation is a real sign of new physics, we will combine  (\ref{eq:DeltaC710}) with~(\ref{eq:deltagLbexp}) to set bounds on the TGC parameters. This exercise will illustrate the power of modes like $B \to K^\ast \mu^+ \mu^-$ in constraining the values of $\Delta g_1^Z$, $\Delta \kappa_\gamma$ and $\lambda_\gamma$. Of course the final results of our analysis should be taken with a pinch of salt, because the anomalies  in the $b \to s \ell^+ \ell^-$ sector may become smaller (or even disappear) once LHCb has analysed the full LHC run-1 data set. 

By building a $\Delta \chi^2$ from (\ref{eq:deltagLbexp}) and (\ref{eq:DeltaC710}), we obtain the allowed 68\% CL ($\Delta \chi^2 = 2.3$) and  95\% CL ($\Delta \chi^2 = 6.0$) regions in the $\Delta g_1^Z \hspace{0.25mm}$--$ \hspace{0.25mm} \Delta \kappa_\gamma$,  $\Delta g_1^Z \hspace{0.25mm}$--$ \hspace{0.25mm} \lambda_\gamma$  and $\lambda_\gamma  \hspace{0.25mm}$--$ \hspace{0.25mm} \Delta \kappa_\gamma$ planes shown in Figure~\ref{fig:panels2}. For $\lambda_\gamma = 0$ (upper left panel), we obtain the 68\% CL constraints 
\beq \label{eq:fit2}
\Delta g_1^Z = -0.003 \pm 0.007 \,, \qquad 
\Delta \kappa_\gamma = 0.13 \pm 0.04 \,, 
\eeq
with the correlations
\beq
\rho = \begin{pmatrix} 1 & -0.32 \\  -0.32 & 1 \end{pmatrix} \,.
\eeq
Comparing the numbers in (\ref{eq:fit2}) with those quoted in  (\ref{eq:fit1}), one sees that including the information on the inclusive and exclusive $b \to s \ell^+ \ell^-$ transitions improves  the constraints that can be put on the TGC parameter $\Delta \kappa_\gamma$ notably. In fact, the uncertainty on $\Delta \kappa_\gamma$ is now comparable to the one found in \cite{Falkowski:2014tna} from a study of $W^+ W^-$ production at LEP~II. Our findings illustrate that precision measurements of $B \to K^{(\ast)} \mu^+ \mu^-$ and $B_s \to \phi \mu^+ \mu^-$, possible at LHCb, provide yet another powerful probe of EW physics. Notice also that the best fit  point in (\ref{eq:fit2}) corresponds to a large destructive photon penguin with $\Delta D \simeq 1$. Such a  photon penguin has to arise from a strongly-coupled theory with new states  not far above the weak scale that predicts  in addition $|C_{\phi B}| \gg |C_{\phi W}|$ and~$C_{\phi B} > 0$. While it seems challenging to construct a viable UV completion with these features, the possibility that an anomalous photon penguin explains (part of) the deviations seen in semi-leptonic $b \to s$ transitions cannot be excluded in a model-independent fashion.\footnote{Needless to say that a photon penguin is unable to address the anomaly observed recently by LHCb in the ratio $R_K$ of $B^+ \to K^+ \mu^+ \mu^-$ and $B^+ \to K^+ e^+ e^-$ branching ratios at low di-lepton invariant mass~\cite{Aaij:2014ora}.}

From the upper right panel, one furthermore observes that in scenarios with $\Delta \kappa_\gamma = 0$ one cannot obtain the results (\ref{eq:fit2}) within the 68\%~CL. This feature is easy to understand by noting that the parameter $\Delta g_1^Z$ enters $\Delta C_9$ relative to $\Delta C_{10}$ with a factor of $-1+4s_w^2 \simeq -0.08$.  A large $Z$-penguin contribution $\Delta C$ that gives rise to $\Delta C_9 =  {\cal O} (-1)$  thus implies $\Delta C_{10} =  {\cal O} (10)$, which is clearly incompatible with data. The lower panel finally illustrates that including $b \to s \ell^+ \ell^-$ transitions in the global fit helps to partly lift the blind direction along $\Delta \kappa_\gamma \simeq 1/3  \, \lambda_\gamma$, because these modes depend on $\Delta C_7$ and $\Delta C_9$, which both are sensitive to~$ \lambda_\gamma$. Still the existing quark-flavour data is unable to probe  values~$|\lambda_\gamma| \lesssim {\cal O} (1)$, which implies, that compared to other collider probes, radiative and rare $b \to s$ modes do not add any relevant  information on  the latter TGC parameter. 

\begin{figure}[!t]
\begin{center}
\includegraphics[width=0.3 \textheight]{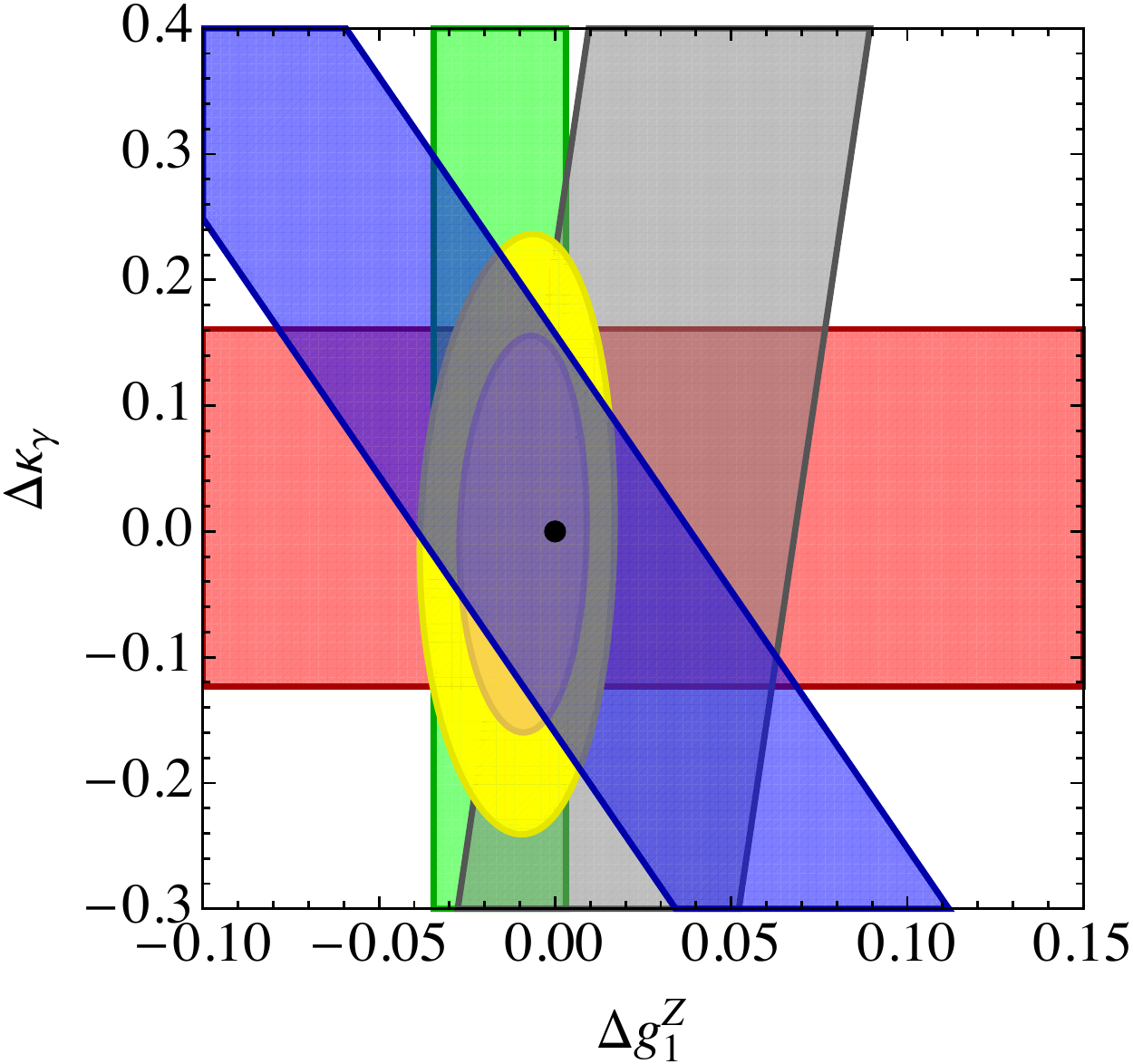}  \qquad 
\includegraphics[width=0.3 \textheight]{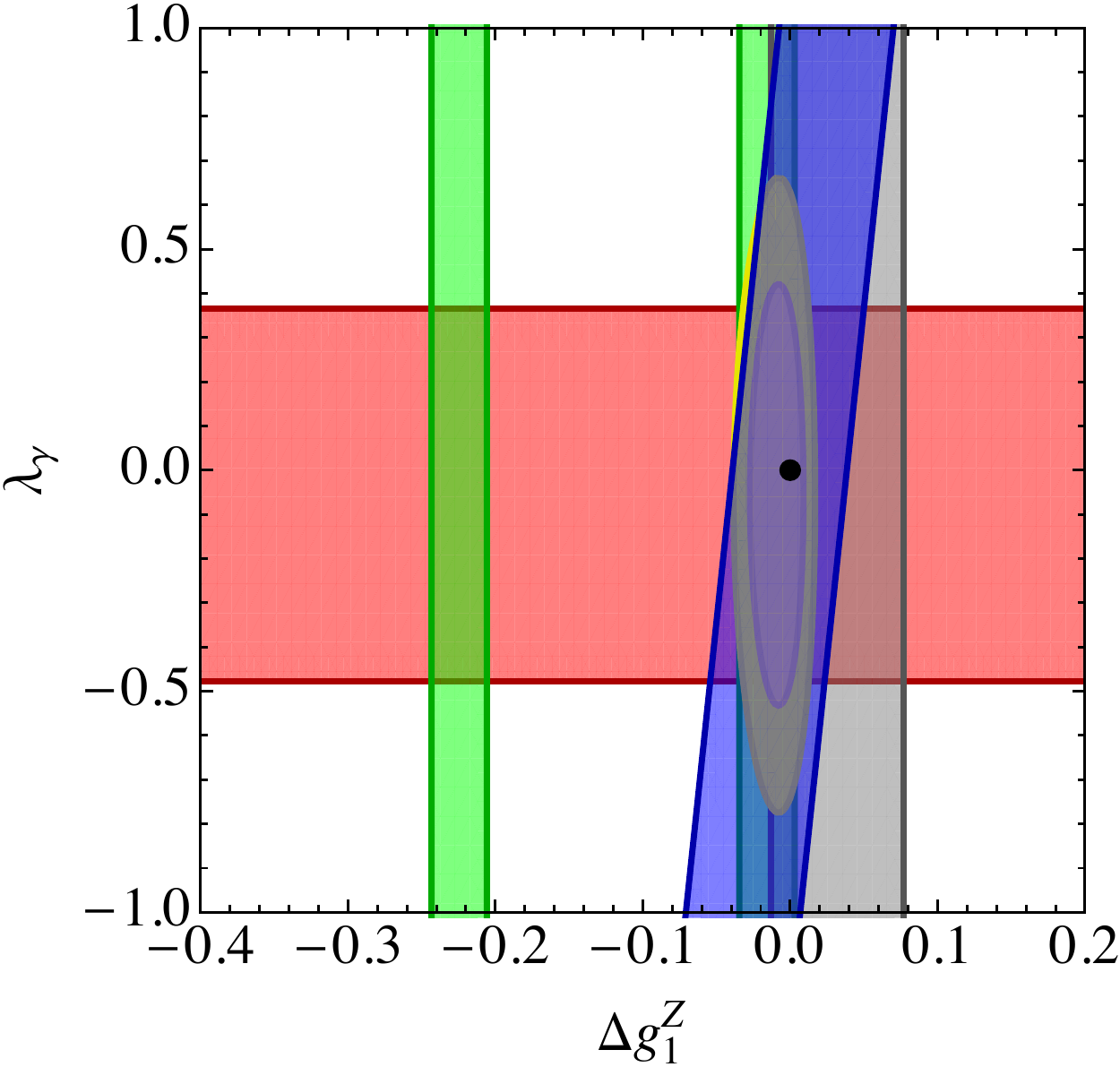} 

\vspace{2mm}

\includegraphics[width=0.3 \textheight]{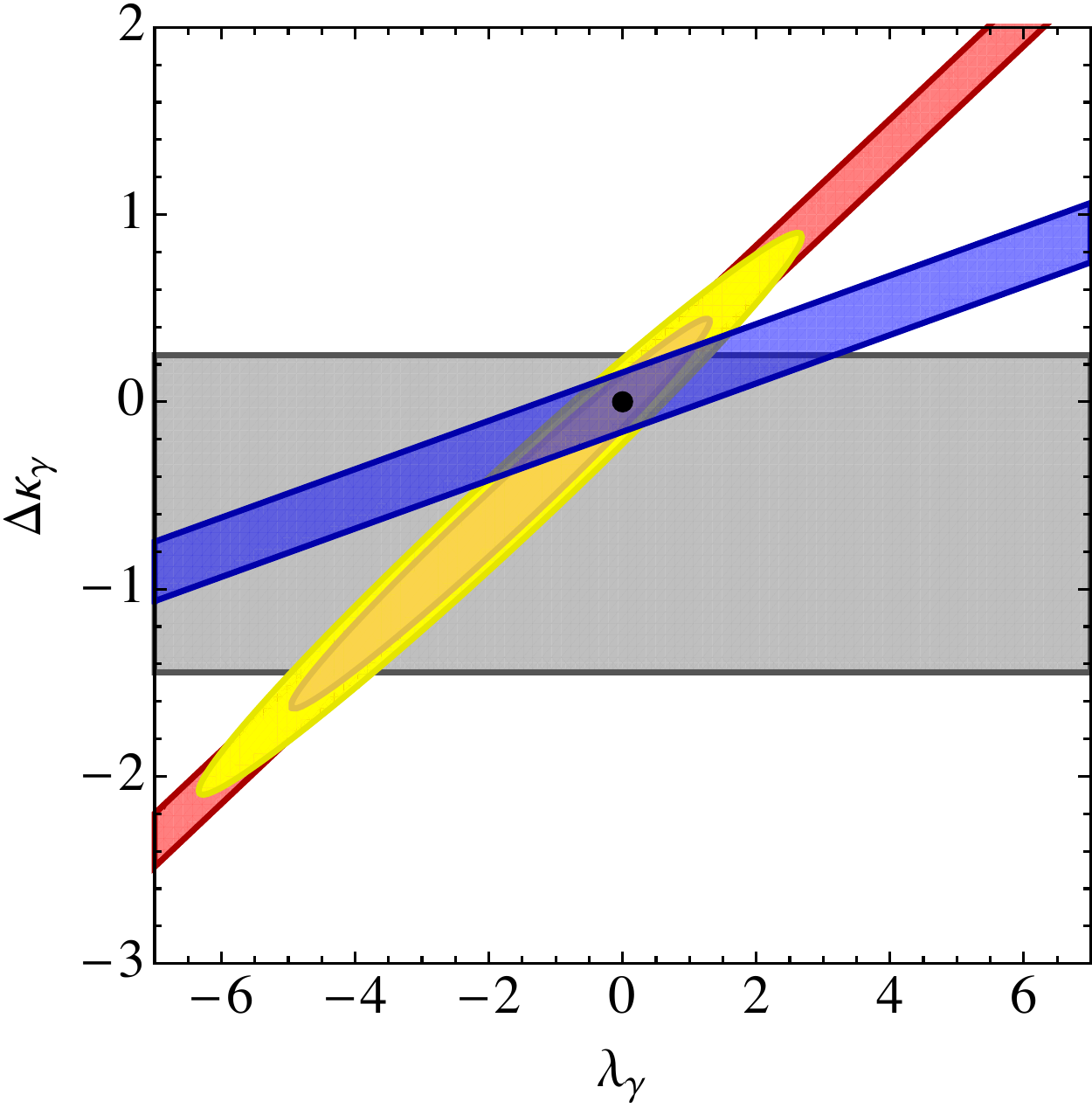} 
\end{center}
\vspace{-4mm}
\caption{\label{fig:panel3}  The blue contours indicate possible future 68\%~CL constraints in the $\Delta g_1^Z \hspace{0.25mm}$--$ \hspace{0.25mm} \Delta \kappa_\gamma$ plane~(upper left panel),  the $\Delta g_1^Z \hspace{0.25mm}$--$ \hspace{0.25mm} \lambda_\gamma$ plane (upper right panel) and the $\lambda_\gamma  \hspace{0.25mm}$--$ \hspace{0.25mm} \Delta \kappa_\gamma$ plane (lower panel)  arising from a future SM calculation of $\epsilon^\prime/\epsilon$ with an assumed total theoretical uncertainty of~$20\%$. The remaining constraints are those shown in Figure~\ref{fig:panels1}.  All  restrictions use $\Lambda = 2 \, {\rm TeV}$.}
\end{figure}

Future precision measurements of the rare $K \to \pi \nu \bar \nu$ decays have the potential to further improve the constraints on the TGC $\Delta g_1^Z$. Employing the results of~\cite{Buras:2005gr, Buras:2006gb, Brod:2008ss, Brod:2010hi}, we obtain 
\beq \label{eq:RKp0}
\begin{split}
R_{K^+} & = \frac{{\rm Br} \left (K^+ \to \pi^+ \nu \bar \nu \right )}{{\rm Br} \left (K^+ \to \pi^+ \nu \bar \nu \right )_{\rm SM}} = 1 +0.93 \hspace{0.25mm} \Delta C  +0.22 \left ( \Delta C \right )^2  \,, \\[2mm]
R_{K^0} & = \frac{{\rm Br} \left (K^0_L \to \pi^0 \nu \bar \nu \right )}{{\rm Br} \left (K^0_L \to \pi^0 \nu \bar \nu \right )_{\rm SM}} = 1 +1.36 \hspace{0.25mm} \Delta C  +0.46 \left ( \Delta C \right )^2  \,. 
\end{split}
\eeq
Using these formulas together with (\ref{eq:DeltaCD}) allows one to translate two-sided limits on $\Delta g_1^Z$ into allowed ranges for $R_{K^+}$ and $R_{K^0}$.  From  (\ref{eq:fit1}), one finds for example 
\beq \label{eq:RKK}
R_{K^+} =[0.81,1.07] \,, \qquad R_{K^0} =[0.73,1.10] \,.
\eeq
These numbers should be compared to the sensitivity that the NA62 \cite{NA62}  and KOTO \cite{KOTO} experiments plan to achieve. The NA62 goal is a measurement of the $K^+ \to \pi^+ \nu \bar \nu$ branching ratio  with 10\% precision, which starts to test the $R_{K^+}$ range quoted in (\ref{eq:RKK}). In contrast, achieving SM sensitivity to the $K^0_L \to \pi^0 \nu \bar \nu$ branching ratio as anticipated by KOTO is  not sufficient to probe the size of the corrections in $R_{K^0}$ that can arise from the presently allowed shifts~$\Delta g_1^Z$.  In order to notably improve the constraints on the TGC~$\Delta g_1^Z$, measurements of the $K \to \pi \nu \bar \nu$ branching ratios at the level of a few percent  thus seem to be needed. In the case of $K^+ \to \pi^+ \nu \bar \nu$, the proposed ORKA experiment \cite{ORKA} may provide such an accuracy. 

Another kaon observable that can provide powerful probes of the EW penguin sector is the ratio $\epsilon^\prime/\epsilon$. Including new-physics effects from both the $Z$-boson  and the photon,  this quantity can be calculated from (see e.g.~\cite{Bauer:2009cf,Buras:2014sba})
\bea \label{eq:epseps}
\begin{split}
 \frac{\epsilon^\prime}{\epsilon} = \Big ( \! \! -2.3 + 21.0   \hspace{0.25mm} R_6 - 8.0   \hspace{0.25mm} R_8  + \Big  [ \left ( 1.2 + 0.2 R_6 \right )  \hspace{0.25mm} \Delta C + \left ( 0.9 - 16.2 \hspace{0.25mm} R_8 \right )  \Delta Z \Big ]  \Big )  \cdot 10^{-4}  \,. \hspace{8mm}
\end{split}
\eea
Here  $R_6$ as well as $R_8$ denote hadronic parameters and 
\beq
\Delta Z = \Delta C + \frac{1}{4} \left (\Delta D - \lambda_\gamma \hspace{0.25mm} f_9^\lambda (x_t) \right ) \,.
\eeq
To obtain useful constraints from (\ref{eq:epseps}), the quantities $R_6$ and $R_8$ have to be known with sufficient precision. While there is at present no reliable result for $R_6$, lattice QCD calculations of  $R_8$ have matured in recent years~\cite{Blum:2012uk} and provide the value $R_8 \simeq 0.83$~\cite{Blum:2015ywa}. Anticipating further progress in the lattice QCD calculations of the hadronic parameters~$R_6$ and  $R_8$,  a full SM calculation of  $\epsilon^\prime/\epsilon$ with a total uncertainty of  20\% may be possible in the near future~(see~Section X.3.2.1 of \cite{Kronfeld:2013uoa}).  

The blue contours in Figure~\ref{fig:panel3} illustrate the impact that a SM calculation of  $\epsilon^\prime/\epsilon$  with a total error of 20\% could have in constraining the TGC parameters $\Delta g_1^Z \hspace{0.25mm}$ and $ \hspace{0.25mm} \Delta \kappa_\gamma$. To obtain the shown 68\%~CL ($\Delta \chi^2 = 2.3$) bounds, we have chosen $R_6= 1.21$ and  $R_8 = 0.83$. With this input our SM prediction reproduces the central value of the experimental determination $\epsilon^\prime/\epsilon = (16.5 \pm 2.6) \cdot 10^{-4}$~\cite{Agashe:2014kda}. From all three panels one observes that with  a theoretical precision of 20\%, the ratio $\epsilon^\prime/\epsilon$ will start to provide  additional meaningful constraints on the TGCs.  Our formula (\ref{eq:epseps}) together with~(\ref{eq:f79}) and (\ref{eq:DeltaCD}) allows to monitor the effect that any theoretical improvement concerning the calculation of the hadronic parameters  $R_6$ and  $R_8$ will have on the extraction of $\Delta g_1^Z$, $\Delta \kappa_\gamma$ and $\lambda_\gamma$ from direct CP violation in the $K \to \pi \pi$ system.

\section{Conclusions}
\label{sec:conclusions}

In this paper we have calculated the one-loop corrections to the flavour-changing electromagnetic dipole and semi-leptonic vector and axial-vector interactions that arise from Feynman diagrams involving anomalous   CP-conserving  TGCs. Although such computations have been performed before by various groups, another  independent calculation was necessary because the analytic results given in the literature differ between publications. With the help of our explicit computation we are able to identify the correct one-loop results and collect them here for the first time. 

Employing our analytic results, we have performed a model-independent EFT study of the constraints on the three TGC parameters $\Delta g_1^Z$, $\Delta \kappa_\gamma$ and $\lambda_\gamma$ using low-energy measurements.  Our analysis is based on the  assumption that the TGC operators are the only SM-EFT interactions that receive non-zero initial conditions at  the new-physics scale $\Lambda$. The magnitude of $\Delta g_1^Z$, $\Delta \kappa_\gamma$ and $\lambda_\gamma$ is then determined from a comparison with the present experimental data on $B \to X_s \gamma$, $B_s \to \mu^+ \mu^-$, $B \to X_s \ell^+ \ell^-$, $B \to K^{(\ast)} \mu^+ \mu^-$, $B_s \to \phi \mu^+ \mu^-$  and $Z \to b \bar b$. We find that while the constraints on the two parameters $\Delta g_1^Z$ and $\Delta \kappa_\gamma$ are strong and competitive with  those obtained from gauge boson pair production at LEP~II, the Tevatron and the LHC, in the case of $\lambda_\gamma$ only weak bounds can be derived. Envisioning experimental progress concerning the measurement of the rare $K \to \pi \nu \bar \nu$ decays as well as theoretical advances in the calculation of the hadronic matrix elements entering  $\epsilon^\prime/\epsilon$, we have furthermore  explored the potential of kaon physics to provide additional informations on the  TGCs. We found that in particular $\epsilon^\prime/\epsilon$ can lead to very valuable  restrictions, because $Z$-boson and photon contributions enter $K \to \pi \pi$ in a different way than for instance $B \to K^\ast \mu^+ \mu^-$. In the future, this feature may help to resolve blind directions in the  TGC parameter space. 

\acknowledgments 
We acknowledge helpful communications with Mikolaj~Misiak and are grateful to  him and  Federico~Mescia for carefully reading  the manuscript and their valuable comments. We thank Joachim~Brod for useful discussions and Veronica~Sanz for encouraging us to write this article. We are grateful to Andrzej Buras for helpful exchange about the hadronic parameter $R_8$ that enters the prediction for~$\epsilon^\prime/\epsilon$, Aneesh~Manohar for explanations concerning his work~\cite{Jenkins:2013wua} and David~Straub for providing further details on his analysis~\cite{Altmannshofer:2014rta}. UH  thanks Michael~Trott for an invitation to the workshop ``HEFT2014 -- Higgs Effective Field Theories''. He also acknowledges the hospitality and support of the CERN theory division.

\end{document}